\documentclass[12pt]{article}
\usepackage[left=2.5cm,top=2.50cm,right=2.5cm,bottom=2.50cm]{geometry}
\usepackage{mathrsfs}
\usepackage{enumitem}
\usepackage{amsmath,amssymb,latexsym,color,cancel,graphicx,bbm,colortbl}
\usepackage[english]{babel}
\usepackage[latin1]{inputenc}
\usepackage{ragged2e}
\usepackage{cite}
\begin{document}
\date{}

\title{Effect of the two-parameter generalized Dunkl derivative on the two-dimensional Schr\"odinger equation}
\author{R. D. Mota$^{a}$, D. Ojeda-Guill\'en$^{b}$\footnote{{\it E-mail address:} dojedag@ipn.mx}}\maketitle

\begin{minipage}{0.9\textwidth}
\small $^{a}$ Escuela Superior de Ingenier{\'i}a Mec\'anica y El\'ectrica, Unidad Culhuac\'an,
Instituto Polit\'ecnico Nacional, Av. Santa Ana No. 1000, Col. San
Francisco Culhuac\'an, Del. Coyoac\'an, C.P. 04430, Ciudad de M\'exico, Mexico.\\

\small $^{b}$ Escuela Superior de C\'omputo, Instituto Polit\'ecnico Nacional,
Av. Juan de Dios B\'atiz esq. Av. Miguel Oth\'on de Mendiz\'abal, Col. Lindavista,
Del. Gustavo A. Madero, C.P. 07738, Ciudad de M\'exico, Mexico.\\

\end{minipage}

\begin{abstract}
We introduce a generalization of the Dunkl-derivative with two parameters to study the Schr\"odinger equation in Cartesian and polar coordinates in two dimensions. The eigenfunctions and the energy spectrum for the harmonic oscillator and the Coulomb problem are derived in an analytical way and it is shown that our results are properly reduced to those previously reported for the Dunkl derivative with a single parameter.

\end{abstract}


\section{Introduction}
The reflection operators were introduced by Wigner \cite{wigner} in the early 50's and applied to the harmonic oscillator by Yang \cite{yang}. These operators have been very useful in the study of the Calogero and Calogero-Sutherland-Moser models \cite{hikami,kakei,lapon}. On the other hand, Dunkl used reflection operators to introduce combined derivative and difference operators. These operators are associated with finite reflection groups and have been very useful to study polynomials in several variables with discrete symmetry \cite{dunkl1,dunkl2}.

Various physical problems involving the Dunkl derivative have been studied by solving the Schr\"odinger equation, including the harmonic oscillator and the Coulomb problem in two and three dimensions \cite{GEN1,GEN2,GEN3,GEN4,nos1,nos2,sami1,sami2}. In references  \cite{GEN1,GEN2,GEN3,GEN4,nos1,nos2,sami1,sami2} the exact solutions of the problems has been found using different analytical and algebraic methods, and properties such as superintegrability have been studied. Similarly, the Dunkl derivative has also been used to study problems in the relativistic regime, by solving both the Klein-Gordon and Dirac equations. Among these relativistic problems are the Coulomb potential, the Klein-Gordon oscillator, and the Dirac-Moshinsky oscillator \cite{nos3,nos4,nos5}.

In addition, various generalizations of the Dunkl derivative have been proposed, which has led to the construction of operators with application in some Calogero-Sutherland models \cite{Chou}. Recently, in references \cite{hassa1,hassa2} the authors introduced a three-parameter Dunkl derivative. With this new generalization it is expected to be able to construct a deformed Schr\"odinger equation that can resolve the discrepancy between theory and experiment.

However, due to its generality, it is not easy to apply this three-parameter Dunkl derivative to physical problems in two or more dimensions. In the present paper we introduce a two-parameter Dunkl derivative that can be applied to physical problems in two and three  dimensions. In particular, we study the harmonic oscillator and the Coulomb potential in two dimensions.

This work is organized as follows. In Section $2$ the two-parameter Dunkl derivative is defined. Then, we give the complete solution of the angular part in terms of the Jacobi polynomials. Section $3$ is dedicated to obtain the radial Schr\"odinger equation for the two-parameter Dunkl derivative for any central potential. In Section $4$, we find the exact solution of the harmonic oscillator in Cartesian and polar coordinates in terms of the generalized Hermite and Laguerre polynomials. Then, we obtain the energy spectrum and eigenfunctions of the Coulomb potential in an analytical way. We show that all our results are adequately reduced to those previously obtained with the standard Dunkl derivative with a single parameter. Finally, we give some concluding remarks.

\section{The Schr\"odinger equation with the two-parameter Dunkl derivative }

We introduce the two-parameter generalized Dunkl derivative defined by
\begin{eqnarray}
&&\tilde D_1\equiv\frac{\partial}{\partial x}+\frac{\mu_1}{x}(1-R_1)+\gamma\frac{\partial}{\partial x}R_1=D_1^{\mu_1}+\gamma\frac{\partial}{\partial x}R_1, \label{dunkl1}\\
&&\tilde D_2\equiv\frac{\partial}{\partial y}+\frac{\mu_2}{y}(1-R_2)+\gamma\frac{\partial}{\partial y}R_2=D_2^{\mu_2}+\gamma\frac{\partial}{\partial y}R_2.\label{dunkl2}
\end{eqnarray}

In these definitions, $D_1^{\mu_1}$and $D_2^{\mu_2}$ are the standard Dunkl derivative in the $x$ ($y$) variable, the constants $\mu_1,\mu_2$ satisfy $\mu_1>-1/2$ and $\mu_2>-1/2$ \cite{GEN4}, and $R_1,R_2$ are the reflection operators with respect to the $x-$ and $y-$ coordinates. Thus, $R_1f(x,y)=f(-x,y)$ and $R_2f(x,y)=f(x,-y)$. The parameter $\gamma$, as it will be clear later on this work, takes the values $-1<\gamma<1 $.

With our definitions (\ref{dunkl1}) and (\ref{dunkl2}), ${\bf P}$ changes to $-i(\tilde D_1, \tilde D_2)$, and ${\bf P}^2=- \nabla^2$ changes to ${\bf P}^2=-\left(\tilde D_1^2+\tilde D_2^2\right)\equiv-\nabla^2_{\tilde D}$, where  $\nabla^2_{\tilde D}$ is the generalized Dunkl-Laplacian. Hence, if we set $\hbar=m=1$, the stationary generalized Schr\"odinger-Dunkl equation takes the form
\begin{equation}
H\Psi\equiv\left(-\frac{1}{2}{\nabla }_{\tilde D}^2+V(x,y)\right)\Psi=E\Psi.\label{DS}
\end{equation}

The action of the reflection operator $R_i$ on a two variables function $f(x,y)$ implies
\begin{equation}
 R_1^2=1, \hspace{3ex}\frac{\partial}{\partial x}R_{1}=-R_1\frac{\partial}{\partial x},\hspace{3ex}R_1x=-xR_1, \hspace{3ex}R_1\tilde D_1=-\tilde D_1R_1, \label{pro1}
\end{equation}
and similar expressions for the $y-$ coordinate. Also, the following equalities involving the operators $R_i$ and $ \tilde D_i$ can be proved
\begin{eqnarray}
&&R_1R_2=R_2R_1,\hspace{5ex}[\tilde D_1, \tilde D_2]=0,\\
&&[x,{\tilde D}_1]=-1+(\gamma -2\mu_1) R_{1}+2\gamma x \frac{\partial}{\partial x}R_{1},\\
&&[y,{\tilde D}_2]=-1+(\gamma -2\mu_2) R_ {2}+2\gamma y \frac{\partial}{\partial y}R_{2}. \label{pro2}
\end{eqnarray}

By direct calculation, we find that the Dunkl-Laplacian in cartesian coordinates takes the form
\begin{eqnarray}
&&\nabla_{\tilde D}^2={ \tilde D}_1^2+{ \tilde D}_2^2\nonumber\\
&&\hspace{4ex}=\left(1-\gamma^2\right)\left(\frac{\partial^2}{\partial x^2}+\frac{\partial^2}{\partial y^2}+2\frac{\frac{\mu_1}{1-\gamma}}{x}\frac{\partial}{\partial x}+2\frac{\frac{\mu_2}{1-\gamma}}{y}\frac{\partial}{\partial y}-\frac{\frac{\mu_1}{1-\gamma}}{x^2}(1-R_1)-\frac{\frac{\mu_2}{1-\gamma}}{y^2}(1-R_2)\right).
\label{DGEN}
\end{eqnarray}
With the definitions
\begin{eqnarray}
&&\eta_1=\frac{\mu_1}{1-\gamma},\hspace{3.4cm} \eta_2=\frac{\mu_2}{1-\gamma},\\\label{etas}
&&D_1^{\eta_{1}}\equiv\frac{\partial}{\partial x}+\frac{\eta_1}{x}(1-R_1), \hspace{1cm} D_2^{\eta_{2}}\equiv\frac{\partial}{\partial y}+\frac{\eta_2}{y}(1-R_{2})\label{ddes}.
\end{eqnarray}
we write the operator of equation (\ref{DGEN}) as
\begin{eqnarray}
&&\nabla_{\tilde D}^2=\left(1-\gamma^2\right)\left((D_1^{\eta_{1}})^2+(D_2^{\eta_{2}})^2\right)\equiv\left(1-\gamma^2\right)\nabla_{\eta_{1}\eta_{2}}^2,
\end{eqnarray}
where $\nabla_{\eta_{1}\eta_{2}}^2$ is the standard Dunkl-Laplacian with the effective parameters $\eta_1$ and $\eta_2$ in the Dunkl derivatives (\ref{ddes}).

In  polar coordinates the generalized Dunkl-Laplacian of expression (\ref{DGEN}) is written as
\begin{equation}
\nabla_{\tilde D}^2= \left(1-\gamma^2\right)\left(\frac{\partial^2}{\partial \rho^2}+\frac{1+2\eta_1+2\eta_2}{\rho}\frac{\partial}{\partial \rho}-\frac{2}{\rho^2}B_\phi \right), \label{laplapol}
\end{equation}
where the operator $B_\phi$ is given by
\begin{equation}
B_\phi\equiv-\frac{1}{2}\frac{\partial^2}{\partial \phi^2}+\left(\eta_1\tan{\phi}-\eta_2\cot{\phi}\right)\frac{\partial}{\partial \phi}
+\frac{\eta_1 (1-R_1)}{2\cos^2{\phi}}+\frac{\eta_2 (1-R_2)}{2\sin^2{\phi}}.
\end{equation}
With these results the stationary generalized Schr\"odinger-Dunkl equation (\ref{DS}) for central potentials takes the form
\begin{equation}
\left(-\frac{1}{2}\left(\frac{\partial^2}{\partial \rho^2}+\frac{1+2\eta_1+2\eta_2}{\rho}\frac{\partial}{\partial \rho}-\frac{2}{\rho^2}B_\phi\right)+\frac{V(\rho)}{1-\gamma^2}\right)\Psi=\frac{E}{1-\gamma^2}\Psi. \label{SD2D}
\end{equation}
It is precisely in this equation where the restriction $-1<\gamma<1$ on the parameter $\gamma$ arises.

From the generalized Dunkl-Laplacian operator (\ref{DGEN}), it is convenient to introduce the Dunkl angular momentum operator
\begin{equation}
{\mathcal J_{\eta_{1}\eta_{2}}}=i(xD_2^{\eta_{2}}-yD_1^{\eta_{1}}),
\end{equation}
which can be used to show the following results
\begin{eqnarray}
&&\left[xD_2^{\eta_{2}},\nabla_{\eta_{1}\eta_{2}}^2 \right]=2D_2^{\eta_{2}}D_1^{\eta_{1}},\label{res1}\\
&&\left[yD_1^{\eta_{1}},\nabla_{\eta_{1}\eta_{2}}^2\right]=2D_1^{\eta_{1}}D_2^{\eta_{2}},\label{res2}\\
&&\left[\frac{\mu_i}{x_i}\left(1-R_i\right),F(\rho)\right]=0, \hspace{2ex}i=1,2,\label{res3}\\
&&\left[\left(x\frac{\partial}{\partial y}-y\frac{\partial}{\partial x}\right),F(\rho)\right]=\left[\frac{\partial }{\partial \phi},F(\rho)\right]=0,\label{res4}
\end{eqnarray}
where $F(\rho)$ is an arbitrary function with partial derivative. In the last two equalities we have used the polar coordinates $\rho=\sqrt{x^2+y^2}$, $\tan{\phi}=\frac{y}{x}$. From these commutation relations, we immediately show that the operator ${\mathcal J}_{\eta_{1}\eta_{2}}$  is a constant of motion of the Hamilton operator $H$
\begin{equation}
[{\mathcal J}_{\eta_{1}\eta_{2}}, H]=0.
\end{equation}
As it will be shown below, this fact will allow us to solve the generalized Schr\"odinger-Dunkl equation (\ref{DS}) by using separation of variables on the wave function.

In polar coordinates the operator ${\mathcal J}_{\eta_{1}\eta_{2}}$ takes the form
\begin{equation}
\mathcal{J}_{\eta_{1}\eta_{2}}=i(\partial_\phi+\eta_2\cot\phi(1-R_2)-\eta_1\tan\phi(1-R_1)),\label{j}
\end{equation}
and therefore, the square of this operator is given by
\begin{equation}
\mathcal{J}^2_{\eta_{1}\eta_{2}}=2B_\phi+2\eta_1\eta_2(1-R_1R_2).\label{jcuad}
\end{equation}

The spectrum and the eigenfunctions of the operator ${\mathcal J}_{\eta_{1}\eta_{2}}$ have been constructed in Ref. \cite{GEN1}. Their construction is based on the fact that the operator $\mathcal{J}_{\eta_{1}\eta_{2}}$ commutes with the operator $R_1R_2$. Thus, they proposed the eigenvalues and eigenvectors in the form
\begin{equation}
\mathcal{J}_{\eta_{1}\eta_{2}}\Phi_\epsilon=\lambda_\epsilon \Phi_\epsilon,\label{angular}
\end{equation}
being $\epsilon\equiv s_1s_2=\pm 1$, and $s_1$, $s_2$ the eigenvalues of the reflection operators $R_1$ and $R_2$, respectively.

In summary, according to Ref. \cite{GEN1} the eigenfunctions and eigenvalues of the operator ${\mathcal J}_{\eta_{1}\eta_{2}}$ are classified in the following two cases:\\
\begin{itemize}
\item If $R_1=R_2$, then $\epsilon=1$. The solutions of equation (\ref{angular}) are
\begin{eqnarray}
&&\Phi_{+} (\phi)=\Phi_\ell^{++}(\phi)\pm i\Phi_\ell^{--}(\phi),\hspace{.5ex}\\
&&\lambda_{+}=\pm2\sqrt{\ell(\ell+\eta_1+\eta_2)},
\end{eqnarray}
where $\ell\in {\mathbb{N}}$, and $\Phi_\ell^{++}$ $\Phi_\ell^{--}$ are given by
\begin{eqnarray}
\Phi_\ell^{++}(x)=\sqrt{\frac{(2\ell+\eta_1+\eta_2)\Gamma{(\ell+\eta_1+\eta_2)\ell !}}{2\Gamma{(\ell+\eta_1+1/2)}\Gamma{(\ell+\eta_2+1/2)}}}P_\ell^{(\eta_1-1/2,\eta_2-1/2)}(x),\hspace{18ex}\label{masmas}\\
\Phi_\ell^{--}(x)=\sqrt{\frac{(2\ell+\eta_1+\eta_2)\Gamma{(\ell+\eta_1+\eta_2+1)(\ell-1)!}}{2\Gamma{(\ell+\eta_1+1/2)}\Gamma{(\ell+\eta_2+1/2)}}}\sin\phi\cos\phi P_{\ell-1}^{(\eta_1+1/2,\eta_2+1/2)}(x).\label{menmen}
\end{eqnarray}
In these expressions $P_\ell^{(\alpha,\beta)} (x)$ are the classical Jacobi polynomials and $x=-\cos2\phi$, such that $P_{-1}^{(\alpha,\beta)}(x)=0$ and as a consequence $\Phi_0^{--}=0$.\\
\item For $R_1=-R_2$, $\epsilon=-1$,
\begin{eqnarray}
&&\Phi_{-} (\phi)=\Phi_\ell^{-+}(\phi)\mp i\Phi_\ell^{+-}(\phi),\hspace{3ex}\\
&&\lambda_{-}=\pm2\sqrt{(\ell+\eta_1)(\ell+\eta_2)},
\end{eqnarray}
where $\ell\in \{\frac{1}{2},\frac{3}{2},...\}$. The expressions for $\Phi_\ell^{-+}$ and $\Phi_\ell^{+-}$ are
\begin{eqnarray}
\Phi_\ell^{-+}(x)=\sqrt{\frac{(2\ell+\eta_1+\eta_2)\Gamma{(\ell+\eta_1+\eta_2+1/2)(\ell-1/2)!}}{2\Gamma{(\ell+\eta_1+1)}\Gamma{(\ell+\eta_2)}}}\cos\phi P_{\ell-1/2}^{(\eta_1+1/2,\eta_2-1/2)}(x),\label{menmas}\\
\Phi_\ell^{+-}(x)=\sqrt{\frac{(2\ell+\eta_1+\eta_2)\Gamma{(\ell+\eta_1+\eta_2+1/2)(\ell-1/2)!}}{2\Gamma{(\ell+\eta_1)}\Gamma{(\ell+\eta_2+1)}}}\sin\phi P_{\ell-1/2}^{(\eta_1-1/2,\eta_2+1/2)}(x).\hspace{.5ex}\label{masmen}
\end{eqnarray}
\end{itemize}
As it will be seen in the next Section, these results will allow us to obtain the radial part of the generalized Schr\"odinger-Dunkl equation for any central potential.

\section{The radial generalized Schr\"odinger equation for the two-parameter Dunkl derivative}

From equation  (\ref{jcuad}) it follows that
\begin{equation}
B_\phi=\frac{1}{2}\left(\mathcal{J}_{\eta_{1}\eta_{2}}^2-2\eta_1\eta_2(1-R_1R_2)\right). \label{bfi}
\end{equation}
Thus, the generalized Schr\"odinger-Dunkl equation (\ref{SD2D}) takes the form
\begin{equation}
\left(-\frac{1}{2}\left(\frac{\partial^2}{\partial \rho^2}+\frac{1+2\eta_1+2\eta_2}{\rho}\frac{\partial}{\partial \rho}-\frac{\mathcal{J}_{\eta_{1}\eta_{2}}^2-2\eta_1\eta_2(1-R_1R_2)}{\rho^2}\right)+\frac{V(\rho)}{1-\gamma^2}\right)\Psi=\frac{E}{1-\gamma^2}\Psi\label{SD2DP}.
\end{equation}
If we propose $\Psi=R(\rho)\Phi(\phi)$, according the results of the preceding Section we have the following cases:\\
\begin{enumerate}[label=(\alph*)]
\item $R_1=R_2$, $\epsilon=s_1s_2=1$. In this case, the centrifugal coefficient of equation (\ref{SD2DP}) reduces to
\begin{equation}
\mathcal{J}_{\eta_{1}\eta_{2}}^2-2\eta_1\eta_2(1-s_1s_2)=\lambda_+^2=4\ell(\ell+\eta_1+\eta_2).\label{eigen1}
\end{equation}
\item $R_1=-R_2$,  $\epsilon=s_1s_2=-1$. Hence, the centrifugal coefficient of equation (\ref{SD2DP}) results to be
\begin{equation}
\mathcal{J}_{\eta_{1}\eta_{2}}^2-2\eta_1\eta_2(1-s_1s_2)=\lambda_-^2-4\eta_1\eta_2=4(\ell+\eta_1)(\ell+\eta_2)-4\eta_1\eta_2.
\label{eigen2}
\end{equation}
\end{enumerate}
Now, the numerical values of equations (\ref{eigen1}) and (\ref{eigen2}) are equal. This implies that the angular part solutions of the generalized Schr\"odinger-Dunkl equation are given according the parities $s_1=s_2$ or $s_1=-s_2$. However, the radial part of the generalized Schr\"odinger-Dunkl equation for both cases is the same, and is given by
\begin{equation}
\left(\frac{d^2}{d \rho^2}+\frac{1+2\eta_1+2\eta_2}{\rho}\frac{d}{d \rho}-\frac{4\ell(\ell+\eta_1+\eta_2)}{\rho^2}-\frac{2V(\rho)}{1-\gamma^2}+\frac{2E}{1-\gamma^2}\right)R(\rho)=0.\label{RadialSD}
\end{equation}
This radial generalized Schr\"odinger-Dunkl equation for any central potential will be solved in the next Section for the harmonic oscillator and the Coulomb problem in two dimensions.

\section{The generalized Schr\"odinger-Dunkl equation for the harmonic oscillator and the Coulomb problem in two dimensions}

\subsection{Analytical solution of the generalized Dunkl-oscillator: Cartesian coordinates}

The generalized Schr\"odinger equation with the two-parameter Dunkl derivative (\ref{DS}) for the potential $V(x,y)=\frac{1}{2}(x^2+y^2)$
is
\begin{equation}
\left\{\sum_{x_i=x,y}\left(\frac{\partial^2}{\partial x_i^2}+\frac{2\eta_i}{x_i}\frac{\partial}{\partial x_i}-\frac{\eta_i}{x_i^2}+\frac{\eta_i}{x_i^2}R_i-\frac{x_i^2}{1-\gamma^2}\right)\right\}\varphi(x,y)=-\frac{2E}{1-\gamma^2}\varphi(x,y).
\end{equation}
By setting $\varphi(x,y)=\varphi_1(x)\varphi_2(y)$ and the separation constants as $-\frac{2E_i}{1-\gamma^2}$  ($i=x,y$) such that $E_1+E_2=E$, we have to solve the following equations
\begin{eqnarray}
&&\left(\frac{d^2}{d x^2}+\frac{2\eta_1}{x}\frac{d}{d x}-\frac{\eta_1}{x^2}+\frac{\eta_1}{x^2}R_1-\frac{x^2}{1-\gamma^2}+\frac{2E_1}{1-\gamma^2}\right)\varphi_1(x)=0,\label{car1}\\
&&\left(\frac{d^2}{d y^2}+\frac{2\eta_2}{y}\frac{d}{d y}-\frac{\eta_2}{y^2}+\frac{\eta_2}{y^2}R_2-\frac{y^2}{1-\gamma^2}+\frac{2E_2}{1-\gamma^2}\right)\varphi_2(y)=0.
\end{eqnarray}
Since the expressions are the same for $x$ and $y$, we will focus our attention on the variable $x$. If we define
\begin{equation}
\tilde{x}(1-\gamma^2)^{1/4}=x,\hspace{6ex}\varepsilon=\frac{2E}{(1-\gamma^2)^{1/2}},
\end{equation}
the equation (\ref{car1}) takes the form
\begin{equation}
\left(\frac{d^2}{d {\tilde x}^2}+\frac{2\eta_1}{ {\tilde x}}\frac{d}{d  {\tilde x}}-\frac{\eta_1}{ {\tilde x}^2}+\frac{\eta_1}{ {\tilde x}^2}R_1-{\tilde x}^2+\varepsilon\right)\varphi_1( {\tilde x})=0.
\end{equation}
We find that the admissible solutions for $s_1=1$ are
\begin{equation}
\varphi^+_1(\tilde x)=\sqrt{\frac{n!}{\Gamma(n+\eta_1+\frac{1}{2})}}e^{-\frac{\tilde x^2}{2}}L_n^{\eta_1-\frac{1}{2}}(\tilde x^2),\label{pares}
\end{equation}
where their corresponding eigenvalues are given by
\begin{equation}
\varepsilon=4n+2\eta_1+1\hspace{3ex}\Rightarrow\hspace{3ex} E_1=\left(2n+\eta_1+1/2\right)\sqrt{1-\gamma^2},\hspace{3ex}n\epsilon\{0,1,2,...\}
\end{equation}
Similarly, for $s_1=-1$ we find that the eigenfunctions and eigenvalues explicitly are
\begin{equation}
\varphi^-_1(\tilde x)=\sqrt{\frac{n!}{\Gamma(n+\eta_2+\frac{3}{2})}}e^{-\frac{\tilde x^2}{2}}\tilde x L_n^{\eta_1+\frac{1}{2}}(\tilde x^2),\label{impares}
\end{equation}
\begin{equation}
\varepsilon=4n+2\eta_1+3\hspace{3ex}\Rightarrow\hspace{3ex} E_1=\left(2n+\eta_1+3/2\right)\sqrt{1-\gamma^2},\hspace{3ex}n\epsilon\{0,1,2,...\}
\end{equation}
Now, we introduce the relation between the generalized Hermite and the Laguerre polynomials \cite{GEN1}
\begin{eqnarray}
&&H_{2n}^\eta(x)=(-1)^n \sqrt{\frac{n!}{\Gamma(n+\eta+\frac{1}{2})}}L_n^{\eta-\frac{1}{2}}(x^2),\\
&&H_{2n+1}^\eta(x)=(-1)^n \sqrt{\frac{n!}{\Gamma(n+\eta+\frac{3}{2})}}\tilde x L_n^{\eta+\frac{1}{2}}(x^2).
\end{eqnarray}
With these relations, we can write the even and odd harmonic oscillator solutions in the compact form
\begin{equation}
\varphi_{n_1}=e^{-\frac{\tilde x^2}{2}} H_{n_1}^{\eta_1} (\tilde x),\hspace{5ex} E_{n_1}=(n_1+\eta_1+1/2)\sqrt{1-\gamma^2},
\end{equation}
where $n_1\epsilon \mathbb{N}$ and its parity corresponds to that of the wave function. Thus, the complete solutions of the two-parameter generalized two-dimensional harmonic oscillator are given by
\begin{eqnarray}
&&\varphi_{n_1n_1}(\tilde x,\tilde y)=e^{-\frac{\tilde x^2}{2}} H_{n_1}^{\eta_1} (\tilde x)e^{-\frac{\tilde y^2}{2}} H_{n_2}^{\eta_2} (\tilde y),\\\label{stotal}
&&E=E_{n_1}+E_{n_2}=(n_1+n_2+\eta_1+\eta_2+1)\sqrt{1-\gamma^2}.
\end{eqnarray}
Moreover, the solutions of equation (\ref{stotal}) satisfy the normalized condition
\begin{equation}
\int_{-\infty}^\infty\int_{-\infty}^\infty \varphi_{n_1n_2}(\tilde x,\tilde y) \varphi^{*}_{n'_1n'_2}(\tilde x,\tilde y)|\tilde x|^{2\eta_1}|\tilde y|^{2\eta_2}d\tilde x d\tilde y=\delta_{n_1n'_1}\delta_{n_2n'_2}.
 \end{equation}
Now, since the parameters $\eta_1$ and $\eta_2$ are given by equation (\ref{etas}), in terms of our original parameters we obtain that the solution of the two-dimensional harmonic oscillator are
\begin{eqnarray}
&&\varphi_{n_1n_2}(\tilde x,\tilde y)=e^{-\frac{\tilde x^2}{2}} H_{n_1}^{\frac{\mu_1}{1-\gamma}} (\tilde x)e^{-\frac{\tilde y^2}{2}} H_{n_2}^{\frac{\mu_2}{1-\gamma}} (\tilde y),\\\label{stotal2}
&&E=\left(n_1+n_2+\frac{\mu_1}{1-\gamma}+\frac{\mu_2}{1-\gamma}+1\right)\sqrt{1-\gamma^2}.
\end{eqnarray}
It is important to note that the eigenfunctions and the energy spectrum that we found are properly reduced to those of the harmonic oscillator with the Dunkl derivative with a single parameter \cite{GEN1}.

\subsection{Analytical solution of the generalized Dunkl-oscillator: polar coordinates}

The generalized Schr\"odinger-Dunkl radial equation for the isotropic harmonic oscillator potential $\frac{1}{2}\rho^2$ is
\begin{equation}
\left(\frac{d^2}{d \rho^2}+\frac{1+2\eta_1+2\eta_2}{\rho}\frac{d}{d \rho}-\frac{4\ell(\ell+\eta_1+\eta_2)}{\rho^2}-\frac{\rho^2}{1-\gamma^2}+\frac{2E}{1-\gamma^2}\right)R(\rho)=0.\label{radialSD}
\end{equation}
If we introduce the change of variable $r=\frac{\rho}{(1-\gamma^2)^{1/4}}$ and the new wave function
\begin{equation}
R(r)=r^{-\frac{1+2\eta_1+2\eta_2}{2}}G(r),\label{RG}
\end{equation}
we can write the equation (\ref{radialSD}) as
\begin{equation}
\left(\frac{d^2}{d r^2}+\frac{2E}{\sqrt{1-\gamma^2}}- r^2+\frac{\frac{1}{4}-(2\ell+\eta_1+\eta_2)^2}{r^2}\right)G(r)=0.\label{nosotros}
\end{equation}
This equation has the same form of the differential equation
 \begin{equation}
u''+\left(4n+2\alpha+2-x^2+\frac{\frac{1}{4}-\alpha^2}{x^2}\right)u=0,\label{levedev}
\end{equation}
which has as solution the functions\cite{LEB}
\begin{equation}
u(x)=C_0e^{-\frac{x^2}{2}}x^{\alpha+\frac{1}{2}}L_n^\alpha(x^2),\hspace{5ex} n=0,1,2,...
\end{equation}
where $C_0$ is a normalization constant. Thus, the comparison between equations (\ref{nosotros}) and (\ref{levedev}) leads to
\begin{eqnarray}
&&G(r)=C_0e^{-\frac{r^2}{2}}r^{\alpha+\frac{1}{2}}L^\alpha_n(r^2)\label{G},\\
&&\alpha=2\ell+\eta_1+\eta_2,\\
&&{E}=(2n+1+2\ell+\eta_1+\eta_2)\sqrt{1-\gamma^2}.\label{uno}
\end{eqnarray}
From equations (\ref{RG}) and (\ref{G}), we obtain the radial functions for the generalized harmonic oscillator
\begin{equation}
R_{n\ell}(r)_0=C_0e^{-\frac{r^2}{2}}r^{2\ell}L_n^{2\ell+\eta_1+\eta_2}(r^2).
\end{equation}
The normalization constant $C_0$ can be determined by using the orthogonality of the Laguerre polynomials
\begin{equation}
\int_0^{\infty}e^{-x}x^{\alpha}\left[L_{n}^{\alpha}(x)\right]^2dx=\frac{\Gamma(n+\alpha+1)}{n!}.\label{norm}
\end{equation}
Thus, from this expression we find that $C_0$ is explicitly given by
\begin{equation}
C_0=\sqrt{\frac{2n!}{\Gamma(n+2\ell+\eta_1+\eta_2+1)}}.
\end{equation}
With this normalization constant we obtain that the normalized eigenfunctions take the form
\begin{equation}
R_{n\ell}(r)_O=\sqrt{\frac{2n!}{\Gamma(n+2\ell+\eta_1+\eta_2+1)}}e^{-\frac{r^2}{2}}r^{2\ell}L_n^{2\ell+\eta_1+\eta_2}(r^2).\label{RRR}
\end{equation}
Therefore, the eigenfunctions and energy spectrum for the $2D$ harmonic oscillator in polar coordinates with the parameters of the generalized Dunkl derivative are explicitly given by
\begin{eqnarray}
&&R_{n\ell}(r)_O=\sqrt{\frac{2n!}{\Gamma(n+2\ell+\frac{\mu_1}{1-\gamma}+\frac{\mu_2}{1-\gamma}+1)}}e^{-\frac{r^2}{2}}r^{2\ell}L_n^{2\ell+\frac{\mu_1}{1-\gamma}+\frac{\mu_2}{1-\gamma}}(r^2),\\
&&E=\left(2n+1+2\ell+\frac{\mu_1}{1-\gamma}+\frac{\mu_2}{1-\gamma}\right)\sqrt{1-\gamma^2}.
\end{eqnarray}
Also, we notice that these radial functions are normalized as in the standard Coulomb problem, according to \cite{GEN1,GEN4}
\begin{equation}
\int_0^{\infty}R_{n\ell}(r)R_{n'\ell}(r){r}^{1+2\eta_1+2\eta_2}d r=\delta_{nn'}.\label{normgen}
\end{equation}
Therefore, as we have pointed out at the end of Section 3, the wavefunctions for the generalized Dunkl-oscillator are $\Psi_{O}=R_{n\ell}(\rho)\Phi_{\pm}(\phi)$ (for $s_1s_2=\pm 1$), and are orthogonal according to the scalar product
\begin{equation}
\langle f,g\rangle=\int_0^{\infty}\int_0^{2\pi}f^{*}(r,\phi)g(r,\phi)|r\cos{\phi}|^{2\eta_1}|r\sin{\phi}|^{2\eta_2}r dr d\phi.\label{prodgen}
\end{equation}
We emphasize that the energy spectrum and the states obtained reduce in full agreement to those found for the Dunkl harmonic oscillator for the standard Dunkl derivative with one parameter \cite{GEN1}.

\subsection{Analytical solution of the generalized Dunkl-Coulomb problem}

Now, we shall study the generalized Schr\"odinger-Dunkl equation for the Coulomb problem. Then, if we consider the potential $V(\rho)=-\frac{k}{\rho}$ for bound states ($E=-|E|=-\mathcal{E}$), the equation (\ref{RadialSD}) takes the form
\begin{equation}
\left(\frac{d^2}{d \rho^2}+\frac{1+2\eta_1+2\eta_2}{\rho}\frac{d}{d \rho}-\frac{4\ell(\ell+\eta_1+\eta_2)}{\rho^2}+\frac{2k}{(1-\gamma^2)\rho}+\frac{2\mathcal{E}}{1-\gamma^2}\right)R(\rho)=0.
\end{equation}
With the definition of the new variable
\begin{equation}
r=2\sqrt{\frac{2\mathcal{E}}{1-\gamma^2}}\rho\equiv \epsilon'\rho,
\end{equation}
this radial equation transforms to
\begin{equation}
\left(r\frac{d^2}{d r^2}+(1+2\eta_1+2\eta_2)\frac{d}{d r}-\frac{4\ell(\ell+\eta_1+\eta_2)}{r}+\frac{2k\epsilon'}{1-\gamma^2}-\frac{r}{4}\right)R(r)=0.\label{coulomb}
\end{equation}

On the other hand, it is known that the differential equation
\begin{equation}
xu''+(\beta+1-2\nu)u'+\left(n+\frac{\beta+1}{2}+\frac{\nu(\nu-\beta)}{x}-\frac{x}{4}\right)u=0,\label{generaleq}
\end{equation}
has as solution the functions \cite{LEB}
\begin{equation}
u(x)=Ce^{-\frac{x}{2}}x^\nu L_n^\beta(x),\hspace{5ex} n=0,1,2,...,\label{colsol}
\end{equation}
being $C$ an arbitrary constant and $L_n^\beta(x)$ the generalized Laguerre polynomials. By comparison of equations (\ref{coulomb}) and (\ref{generaleq}), we obtain the following equations
\begin{equation}
\beta-2\nu=2\eta_1+\eta_2,\hspace{7ex}
\nu(\nu-\beta)=-4\ell(\ell+\eta_1+\eta_2),\hspace{7ex}
n+\frac{\beta+1}{2}=\frac{2k\epsilon'}{1-\gamma^2}.  \label{rest}
\end{equation}
From the first two equations we find
\begin{eqnarray}
&&\nu=2\ell,\\ \label{lagc1}
&&\beta=4\ell+2\eta_1+2\eta_2.\label{lagc2}
\end{eqnarray}
We obtain the energy spectrum from the last of equations (\ref{rest})
\begin{equation}
E=-\,{\frac {{2k}^{2}}{ \left( 2n+4\ell+2\eta_1+2\eta_2+1 \right) ^{2}
 \left( 1-\gamma^2 \right)}}.\label{coulespectro}
\end{equation}
Thus, the radial eigenfunctions for our problem are given by
\begin{equation}
R(r)=Ce^{-\frac{r}{2}}r^{2\ell} L_n^{4\ell+2\eta_1+2\eta_2}(r)=Ce^{-\frac{\epsilon'\rho}{2}}(\epsilon'\rho)^{2\ell} L_n^{4\ell+2\eta_1+2\eta_2}(\epsilon'\rho).\label{radial}
\end{equation}
The normalization constant $C$ for this problem is obtained from the integral relationship
\begin{equation}
\int_0^{\infty}e^{-x}x^{\alpha+1}\left[L_{n}^{\alpha}(x)\right]^2dx=\frac{\Gamma(n+\alpha+1)}{n!}(2n+\alpha+1),\label{norm+1}
\end{equation}
which leads us to
\begin{equation}
C=\sqrt{\frac{n!(\epsilon')^{2\eta_1+2\eta_2+2}}{\Gamma(n+4\ell+2\eta_1+2\eta_2+1)(2n+4\ell+2\eta_1+2\eta_2+1)}},
\end{equation}
Therefore, the normalized radial functions for the Coulomb problem of the generalized Schr\"odinger-Dunkl equation are given by
\begin{equation}
R_{n\ell}(\rho)=\sqrt{\frac{n!(\epsilon')^{2\eta_1+2\eta_2+2}}{\Gamma(n+4\ell+2\eta_1+2\eta_2+1)(2n+4\ell+2\eta_1+2\eta_2+1)}}e^{-\frac{\epsilon'\rho}{2}}(\epsilon'\rho)^{2\ell} L_n^{4\ell+2\eta_1+2\eta_2}(\epsilon'\rho),\label{colsolfin}
\end{equation}
where $n=0,1,2,...$.

We must keep in mind that in the equations (\ref{coulespectro}) and (\ref{colsolfin}) the parameters $\eta_1$ and $\eta_2$ are given by equation (\ref{etas}).

Also, in this case the radial functions are normalized as in the standard Coulomb problem \cite{GEN1,GEN4} according to equation (\ref{normgen}). The complete wavefunctions for the generalized Dunkl-Coulomb problem are $\Psi_{C}=R_{n\ell}(\rho)\Phi_{\pm}(\phi)$ (for $s_1s_2=\pm 1$) and are orthogonal according to the scalar product (\ref{prodgen}). Moreover, the energy spectrum and the states obtained are properly reduced to those found for the Dunkl-Coulomb problem with the standard one-parameter Dunkl derivative \cite{GEN4}.

\section{Concluding Remarks}

The generalization of the Dunkl derivative with several parameters was introduced to try to fit the theory with the experiments. In this paper we have introduced a generalization of the Dunkl derivative with two parameters and used it to study the generalized Schr\"odinger equation in two dimensions. In particular, we obtained the energy spectrum and eigenfunctions of the harmonic oscillator in Cartesian and polar coordinates and the Coulomb potential in terms of the generalized Hermite and Laguerre polynomials.

It should be noted that our definition allows us to substitute the two-parameter Dunkl derivative into the Schr\"odinger equation and apply it to solve other important physical problems in $2D$ and $3D$ dimensions, such as Landau levels, the anharmonic oscillator, the Mie-type potential, among others. Moreover, the two-parameter Dunkl derivative defined in the present paper can be applied to study problems in $n$ dimensions, since the eigenfunctions of the Dunkl-Laplace operator have been constructed in Ref. \cite{sami3}.

On the other hand, it is important to point out that as far as we know, no connection has been found between the parameters of the Dunkl derivative and any physical experiment. However, the Dunkl derivative and its different generalizations are currently a relevant field of study in different branches of physics, as can be seen in the Refs \cite{hassa3,Hamil1,Schulze1,Najafizade1,Hamil2,Merad,Dong1,Dong2,Najafizade2,Schulze2,Chung,Nabulsi}.

\section*{Acknowledgments}
This work was partially supported by SNI-M\'exico, COFAA-IPN, EDI-IPN, EDD-IPN, and CGPI-IPN Project Number 20220405.

\end{document}